\documentclass[aps, prl, twocolumn, superscriptaddress, floatfix]{revtex4-2}
\usepackage{graphicx}
\usepackage{amsmath}
\usepackage{ulem}
\usepackage{amssymb}
\usepackage{tikz}
\usepackage{marvosym}
\usepackage{braket}
\usepackage{float}
\usepackage{pgfplots}
\usepackage{mathrsfs}
\usepackage[version=4]{mhchem}
\pgfplotsset{compat=1.17}
\usepackage{subcaption}
\usetikzlibrary{shapes.geometric}
\usetikzlibrary{shadows,patterns,perspective}
\usetikzlibrary {decorations,decorations.text}
\usetikzlibrary{decorations.pathreplacing}
\usepackage{float}
\usepackage{hyperref}
\usepackage{xcolor}

\definecolor{linkcolor}{RGB}{0, 0, 255}      
\definecolor{citecolor}{RGB}{0, 128, 0}     
\definecolor{urlcolor}{RGB}{255, 0, 0}       

\hypersetup{
    colorlinks=true,
    linkcolor=linkcolor,
    citecolor=citecolor,
    urlcolor=urlcolor,
    linktoc=all,  
    pdfborder={0 0 0}  
}

\DeclareCaptionJustification{justified}{\leftskip=0pt \rightskip=0pt \parfillskip=0pt plus 1fil}

\begin{document}
\definecolor{dy}{rgb}{0.9,0.9,0.4}
\definecolor{dr}{rgb}{0.95,0.65,0.55}
\definecolor{db}{rgb}{0.5,0.8,0.9}
\definecolor{dg}{rgb}{0.2,0.9,0.6}
\definecolor{BrickRed}{rgb}{0.8,0.3,0.3}
\definecolor{Navy}{rgb}{0.2,0.2,0.6}
\definecolor{DarkGreen}{rgb}{0.1,0.4,0.1}

\title{What is the topological dual of the XXZ spin Chain?}

\author{Yicheng Tang}
\email{tang.yicheng@rutgers.edu}
\affiliation{Department of Physics and Astronomy, Center for Material Theory, Rutgers University,
Piscataway,  New Jersey, 08854, United States of America}
\author{Pradip Kattel}

\affiliation{Department of Physics and Astronomy, Center for Material Theory, Rutgers University,
Piscataway, New Jersey, 08854, United States of America}

\author{Natan Andrei}
\affiliation{Department of Physics and Astronomy, Center for Material Theory, Rutgers University,
Piscataway,  New Jersey, 08854, United States of America}

\begin{abstract}
We construct a dual symmetry-protected topological (SPT) Hamiltonian for the $U(1)$ symmetric anisotropic spin-$\frac{1}{2}$ Heisenberg chain—a model that has traditionally been used to study spontaneous symmetry breaking (SSB) in both ferromagnetic and antiferromagnetic phases, with an intervening extended Luttinger liquid phase. By performing a non-local unitary transformation, we explicitly construct a local fermionic Hamiltonian that exhibits two nontrivial topological phases separated by an extended Luttinger liquid regime. We demonstrate the topological nature of these phases by analyzing the entanglement structure, deriving a non-local string order parameter, and constructing an exact zero mode operator that connects states in different fermionic parity sectors.
\end{abstract}

\maketitle
Over the past several decades, many studies have established deep connections between seemingly distinct quantum phases. Conventional phases are typically characterized through spontaneous symmetry breaking (SSB) within the Landau-Ginzburg framework~\cite{landau1937theory,Ginzburg:1950sr}, where long-range order is identified by local correlation functions exhibiting a finite expectation value at large distances. In contrast, symmetry-protected topological (SPT) phases do not exhibit conventional order detectable by local correlations~\cite{pollmann2012symmetry}. Instead, they are characterized by non-local string order parameters while also featuring robust edge modes and distinctive entanglement properties associated with long-range entanglement~\cite{chen2011classification}. Foundational work by Tasaki and Kennedy demonstrated  however that nonlocal transformations can bridge the gap between symmetry-protected topological (SPT) phases and phases with spontaneous symmetry breaking (SSB)~\cite{kennedy1992hidden}. In particular, they rigorously established that the topological Haldane chain—now understood as an SPT phase protected by a $\mathbb{Z}_2 \times \mathbb{Z}_2$ symmetry—can be mapped via a nonlocal unitary transformation, now known as the Kennedy–Tasaki (KT) transformation, to a phase in which this $\mathbb{Z}_2 \times \mathbb{Z}_2$ symmetry is spontaneously broken~\cite{kennedy1992hidden1}. Oshikawa later provided an explicit construction of this transformation and extended it to spin chains with any integer spin~\cite{oshikawa1992hidden}. Another explicit example of this correspondence is provided by the one-dimensional spin-$\frac{1}{2}$ cluster-state Hamiltonian, which realizes a nontrivial $\mathbb{Z}_2 \times \mathbb{Z}_2$ SPT phase~\cite{li2023noninvertible}. Under the KT transformation, this model is exactly mapped onto two decoupled transverse-field Ising models, each of which exhibits conventional $\mathbb{Z}_2$ symmetry breaking. A similar paradigm is found in the mapping of the transverse-field Ising chain onto the Kitaev chain via the Jordan–Wigner transformation, where a chain with a spontaneously broken $\mathbb{Z}_2$ symmetry is revealed to have an underlying fermionic structure with topological order~\cite{kitaev2001unpaired}.

The examples above underscore the powerful role of nonlocal unitary transformations in unveiling deep connections between symmetry-protected topological (SPT) order and conventional spontaneous symmetry breaking (SSB) in one-dimensional systems. In this context, the anisotropic Heisenberg chain, historically known as the XXZ model, has long served as a paradigmatic system for exploring spontaneous symmetry breaking. Notably, Orbach solved this model in 1958~\cite{orbach1958linear} exactly using the Bethe Ansatz method~\cite{bethe1931theorie}, establishing its central role in the study of symmetry-breaking phenomena. 
The boundary effects of the spin‑$\frac{1}{2}$ XXZ chain with open boundary conditions have been studied~\cite{pasnoori2023spin,kattel2024edge,PhysRevB.111.L220402,zvyagin2021majorana,zvyagin2024strong}. Yet, despite its importance, an explicit construction of a local dual model that exhibits symmetry-protected topology has yet to be explicitly constructed. In this work, we do just that \footnote{ We note that the  XYZ chain and its SPT dual have been extensively analyzed~\cite{fendley2016strong,katsura2015exact,PhysRevB.105.115406}. However, taking their XXZ limit recovers precisely the trivial Hamiltonian of Eq.\,\eqref{trivham} with no topological phases. In contrast, by employing a modified Jordan–Wigner transform (Eq.\,\eqref{nlt}), our construction maps the XXZ limit onto the dual topological Hamiltonian (Eq.\,\eqref{modelham}) and unveils an extended critical regime separating two distinct SPT phases—features and parameter regimes that cannot be accessed in any limiting case of prior studies.}.

The XXZ Hamiltonian is given by  
\begin{equation}
    H_{\rm XXZ} = \sum_{i=1}^{N-1} \left( \sigma^x_i \sigma^x_{i+1} + \sigma^y_i \sigma^y_{i+1} + \Delta\, \sigma^z_i \sigma^z_{i+1} \right)
    \label{TheXXZham}
\end{equation}

The well known phase diagram of the model consists of two spontaneously symmetry-broken phases (ferromagnetic and antiferromagnetic) separated by an extended critical Luttinger liquid phase~\cite{giamarchi2003quantum}. We will construct a non-local unitary transformation that maps the model onto a local Hamiltonian that realizes two distinct symmetry-protected topological (SPT) phases, similarly separated by an extended critical region

Usually, the XXZ chain is mapped to a Fermion chain via Jordan-Wigner transformation $f^\dagger_i = \frac{1}{2}(\prod_{j<i} \sigma^z_j) (\sigma^x_i+i\sigma^y_i)$, where the $U(1)$ symmetry is preserved to obtain  a Fermionic Hamiltonian of the form
\begin{equation}
    H_f = \sum_{i=1}^{N-1}2(f^\dagger_{i+1}f_i+ f^\dagger_{i}f_{i+1})+\Delta (2n^f_i-1)(2n^f_{i+1}-1)
    \label{trivham}
\end{equation}
where $n^f = f^\dagger f$ and for $\Delta>1$, the model is in a gapped phase that exhibits a spontaneously broken-symmetry ground state with
a charge density wave, for $\Delta<-1$, the model is in the gapped Mott insulating phase, and for $|\Delta|<1$, the model is in the critical phase described by Luttinger liquid. 

However, we may map the XXZ model to another fermionic model, which exhibits topological properties.  Performing the Jordan–Wigner transformation in a slightly different way (see Appendix \ref{dets}), 
\begin{equation}
    c^\dagger_k = \frac{1}{2}(\prod_{j<k} \sigma^x_j)(\sigma^y_k+i\sigma^z_k),
    \label{nlt}
\end{equation}
we obtain a Fermionic model that does not conserve Fermion number but conserves Fermionic parity,
\begin{align}
    &H=\sum_{i=1}^{N-1}(1+\Delta)(c^\dagger_{i+1}c_i+c_i^\dagger c_{i+1} )\nonumber\\
    &+(1-\Delta)(c_ic_{i+1}+c_{i+1}^\dagger c_i^\dagger)+(2n^c_i-1)(2n^c_{i+1}-1),
    \label{modelham}
\end{align}
where  $n_j^c = c_j^\dagger c_j$. We shall show that the model is in a topological phase when $|\Delta| > 1$ and in a topologically trivial critical phase when $|\Delta| < 1$. Moreover, the two topological phases corresponding to $\Delta > 1$ and $\Delta < -1$ are distinct.

It is important to stress that the two models Eq.\eqref{trivham} and Eq.\eqref{modelham} have the same spectrum, which can be obtained exactly using Bethe Ansatz and are related to one another by a non-local unitary transformation such that the two topologically trivial gapped phases of Hamiltonian Eq.\eqref{trivham} becomes topological in Eq.\eqref{modelham}. Notice that the transformation from the $c$ fermion to $f$ fermion is a non-local unitary transformation of the form
\begin{align}
    c_k^\dagger&=\frac{i}{2}
\biggl[\prod_{j=1}^{k-1}(1-2n^f_j)^{k-1-j}\biggr]
\biggl[\prod_{l=1}^{k-1}(f_l+f_l^\dagger)\biggr]\nonumber\\
&\times\bigl[(2n^f_k-1)-\bigl(\prod_{j=1}^{k-1}(1-2n^f_j)\bigr)(f_k^\dagger-f_k)\bigr].
\end{align}

\begin{figure}
    \centering
    \includegraphics[width=\linewidth]{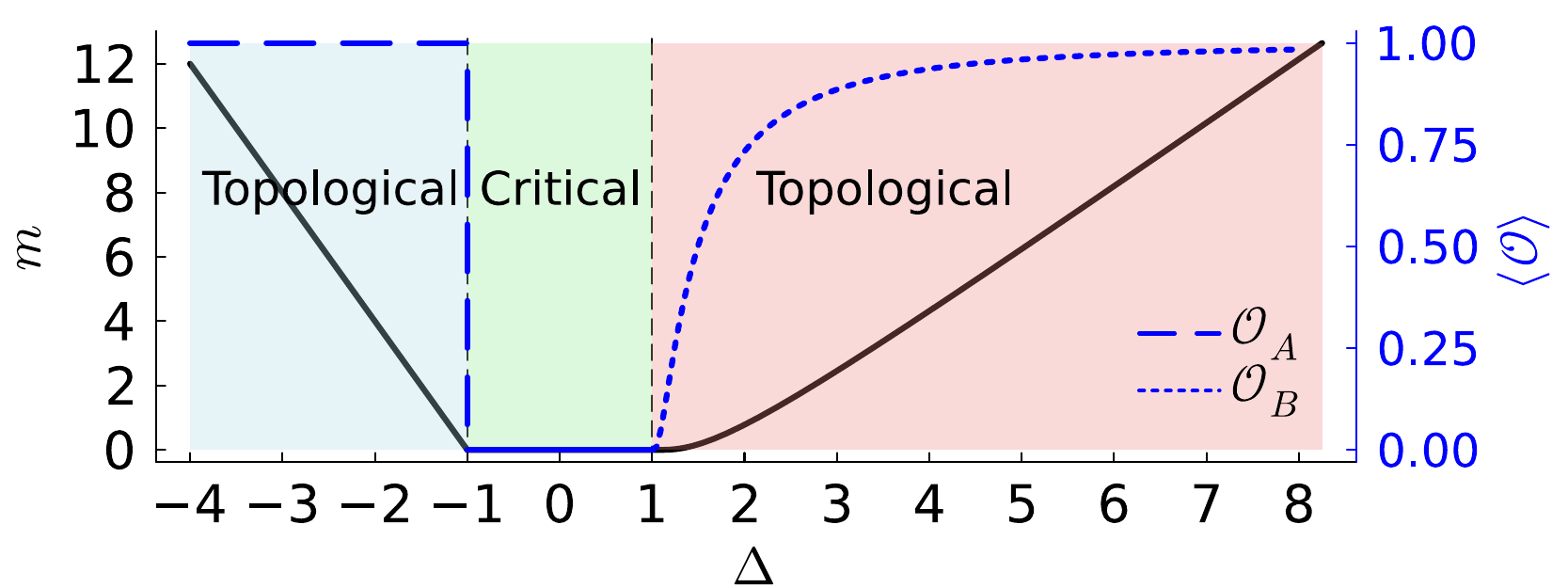}
    \caption{Phase diagram of the Hamiltonian Eq.\eqref{modelham} which is topological when $|\Delta|>1$ and has extended critical phase when $|\Delta|<1$. The black solid line shows the gap, and the blue dashed and blue dotted lines show the expectation value of non-local order parameters (Eq.\eqref{OAdef} and Eq.\eqref{OBdef}) in the parametric regimes $\Delta<-1$ and $\Delta>1$, respectively, both of which are obtained exactly in the thermodynamic limit via Bethe Ansatz.}
    \label{fig:pd}
\end{figure}
\begin{figure}
    \centering
\includegraphics[width=1\linewidth]{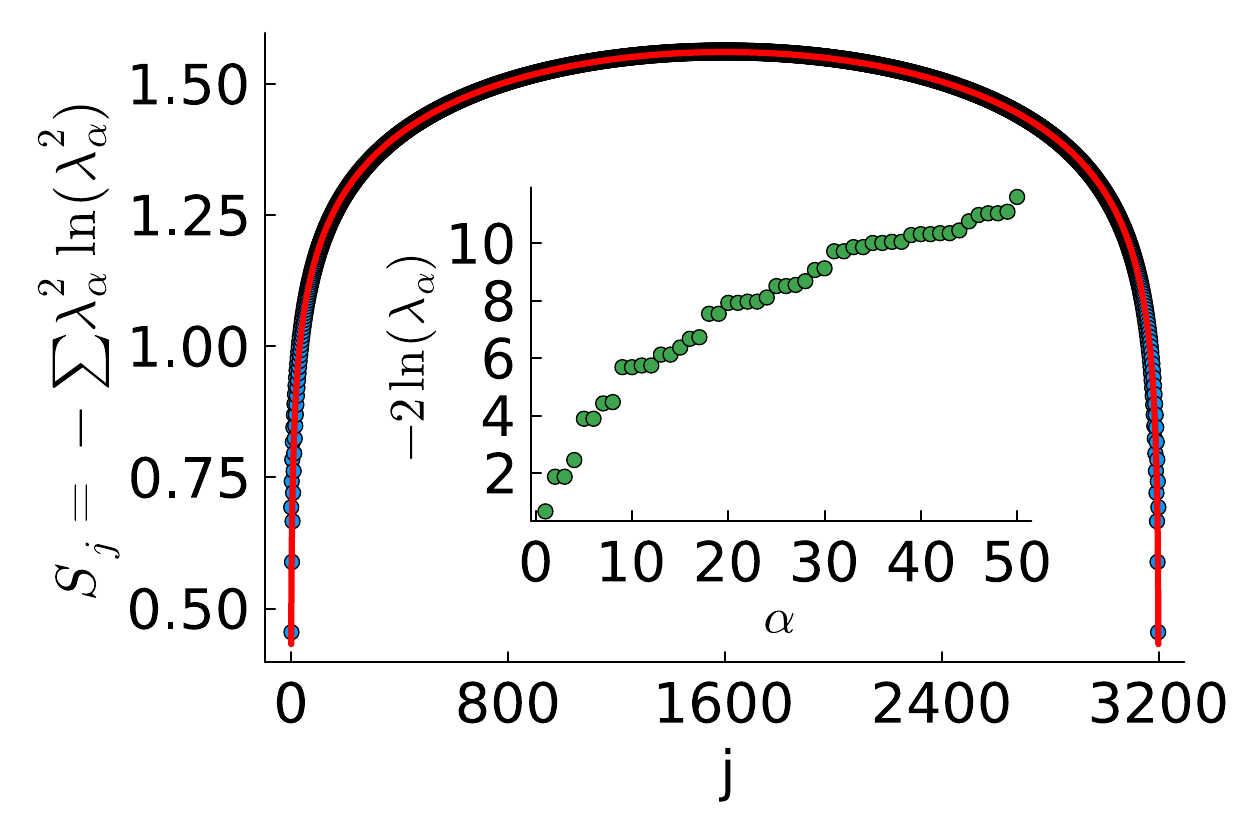}
    \caption{For $|\Delta|<1$, the entanglement spectrum of both Eq.~\eqref{trivham} and Eq.~\eqref{modelham} coincide. Accordingly, we compute the entanglement entropy at anisotropy $\Delta=0.5$ for a spin chain of length $N=3200$ via DMRG, performing 55 sweeps with an SVD truncation threshold of $10^{-10}$. The calculation reaches a maximum bond dimension of 645. The resulting entropy profile is well described by the Cardy–Calabrese formula,
$
S(j)=\frac{c}{6}\ln\!\Biggl[\frac{2N}{\pi a}\sin\!\Bigl(\frac{\pi j}{N}\Bigr)\Biggr],
$
from which we extract a central charge $c=0.99\approx1$. In the inset of Fig.~X, the lowest 50 entanglement levels are shown, defined by
$
E_\alpha=-2\ln\lambda_\alpha,
$
where $\lambda_\alpha^2$ are the eigenvalues of the reduced density matrix obtained by tracing out half of the chain.
}
    \label{fig:delc}
\end{figure}

In the regime $|\Delta|°<1$, as mentioned earlier, the system lies in a gapless phase whose infrared behavior is described by Luttinger liquid theory with central charge $c=1$   ~\cite{haldane1981luttinger,voit1995one,giamarchi2003quantum}.  The entanglement spectrum does not exhibit even degeneracy throughout the spectrum. In Fig~\ref{fig:delc}, we display the entanglement entropy and the associated spectrum for $\Delta = 0.5$, computed using the DMRG algorithm implemented in the ITensor library. Using the Cardy-Calabrese formula $S(j) = \frac{c}{6} \ln\left[ \frac{2L}{\pi a} \sin\left( \frac{\pi j}{N} \right) \right]$, we extract the central charge to be $c=1$ by fitting the entanglement entropy obtained from $N=3200$ sites lattice. 

We emphasize that in the Luttinger liquid phase, the entanglement structure—specifically, the entanglement entropy $S_j$ and the half-chain entanglement spectrum—are independent of the choice of basis. In particular, for both Hamiltonians in Eqs.~\eqref{trivham} and~\eqref{modelham}, the entanglement features are identical, as shown in Fig.~\ref{fig:delc}. 
In contrast, as we shall see, in the two gapped phases the entanglement structure depends explicitly on the basis, reflecting the underlying topological distinctions between the Hamiltonians.

 We now turn to display the topological nature of the two gapped phases when  $|\Delta| >1$. In the regime $\Delta<-1$, it is convenient to consider the Hamiltonian in the Majorana fermion basis. Introducing, $ c_j = \frac{1}{2}(\gamma_{2j-1} + i\gamma_{2j}) $
and
$ c_j^\dagger = \frac{1}{2}(\gamma_{2j-1} - i\gamma_{2j}) $, the Hamiltonian can be written as
\begin{align}
    H &=  \sum_{j=1}^{N-1} -i\Delta\gamma_{2j} \gamma_{2j+1}  \nonumber\\
    &+ \sum_{j=1}^{N-1} i  \gamma_{2j-1} \gamma_{2j+2}- \gamma_{2j-1} \gamma_{2j} \gamma_{2j+1} \gamma_{2j+2}.
    \label{hammajorana}
\end{align}

To compute the exact ground state energy and wavefunction of this interacting Hamiltonian, we define complex fermions from Majorana pairs as
$
\psi_{2j,2j+1} = \frac{1}{2} \Bigl(\gamma_{2j} + i\,\gamma_{2j+1}\Bigr),
$ and show that the ground state takes the form
$
|\mathrm{GS}\rangle = \bigotimes_{j=1}^{N} \left| n_{\psi_{2j,2j+1}}=0 \right\rangle,
$
which satisfies
$
-i\,\gamma_{2j}\gamma_{2j+1} |\mathrm{GS}\rangle = |\mathrm{GS}\rangle
$. The first term in Eq.\eqref{hammajorana} gives an energy $\Delta$ per dimer via the above relation, while the second and third terms acting on the ground state $ |\mathrm{GS}\rangle $ exactly cancel
such that
$
H\,|\mathrm{GS}\rangle = \Delta\,(N-1)\,|\mathrm{GS}\rangle,
$
which shows that  in the $ |\mathrm{GS}\rangle $ there are two unpaired Majorana modes at $\gamma_1$ and $\gamma_{2N}$ encoding a fermionic parity qubit.

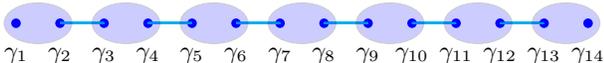
\begin{figure}[H]
    \centering
\begin{tikzpicture}
       \begin{scope}[scale=0.9]
        \def\n{14} 
    \def\spacing{0.65}
    \foreach \i in {0,...,13} {
        \ifnum\i<2
            \fill[blue] (\i*\spacing, 0) circle (2pt);
        \else
            \fill[blue] (\i*\spacing, 0) circle (2pt);
        \fi
        \foreach \i in {1,3,...,11} {
    \draw[thick,cyan] (\i*\spacing, 0) -- ({(\i+1)*\spacing}, 0);
}

        \node[below, yshift=-6pt] at (\i*\spacing, 0) {$\gamma_{\the\numexpr+1+\i\relax}$};
    }
    \draw[fill=blue, opacity=0.2] (0.5*\spacing, 0) ellipse (0.5 and 0.3);
    \draw[fill=blue, opacity=0.2] (2.5*\spacing, 0) ellipse (0.5 and 0.3);
    \draw[fill=blue, opacity=0.2] (4.5*\spacing, 0) ellipse (0.5 and 0.3);
    \draw[fill=blue, opacity=0.2] (6.5*\spacing, 0) ellipse (0.5 and 0.3);
    \draw[fill=blue, opacity=0.2] (8.5*\spacing, 0) ellipse (0.5 and 0.3);
    \draw[fill=blue, opacity=0.2] (10.5*\spacing, 0) ellipse (0.5 and 0.3);
    \draw[fill=blue, opacity=0.2] (12.5*\spacing, 0) ellipse (0.5 and 0.3);
   \end{scope}

\end{tikzpicture}
    \caption{Pictorial representation of the exact ground state of Hamiltonian Eq.\eqref{modelham} in the $\Delta<-1$ regime for a chain of length $N=7$ with two decoupled Majorana fermions as the weak zero mode.}
    \label{fig:enter-label}
\end{figure}

Moreover, bipartitioning the chain at a dimer yields a Schmidt decomposition
$
|\mathrm{GS}\rangle = \frac{1}{\sqrt{2}}\, |\Phi_1^L\rangle \otimes |\Phi_1^R\rangle
+ \frac{1}{\sqrt{2}}\, |\Phi_2^L\rangle \otimes |\Phi_2^R\rangle,
$
with Schmidt coefficients $\lambda_1 = \lambda_2 = 1/\sqrt{2}$ and $\ket{\Phi}$s are the Schmidt vector. Consequently, the entanglement entropy  for any cut is
$
S = -\sum_{\alpha=1}^2 \lambda_\alpha^2 \ln \lambda_\alpha^2 = \ln2,
$
and the entanglement spectra are exactly doubly degenerate with $\epsilon_\alpha=-2\log\lambda_\alpha=\ln(2)$ for $\alpha=1,2$
which reflects the topological degeneracy and the protected edge modes characteristic of a one-dimensional topological superconductor~\cite{kitaev2001unpaired}.

Since the two non-vanishing entanglement spectra values are exactly degenerate, the model is in the topological phase. This can further be verified by the explicit construction of the non-local string order parameter 
\begin{equation}
    \mathcal{O}_B^2=\lim_{|i-j|\to \infty}(c_i^\dagger + c_i)\,\exp\left(i\pi\sum_{n=i}^{j-1} c_n^\dagger c_n\right)\,(c_j^\dagger + c_j),
\end{equation}
whose explicit computation is simpler in the spin basis upon performing the inverse transformation Eq.\eqref{nlt} such that
\begin{equation}
    \mathcal{O}_B=\sqrt{\langle \mathcal O^2_B \rangle}=\lim_{|i-j|\to \infty}\sqrt{\langle \sigma_i^z \sigma_j^z \rangle} = 1
    \label{OBdef}
\end{equation}
which as shown in the phase diagram Fig.\eqref{fig:pd} is constant for the entire regime $\Delta<-1$. Moreover, we can also explicitly construct the Majorana operator connecting the even and odd fermionic parity ground state as $M_1=c_1+c_1^\dagger$, which is an exact weak zero mode that connects the two degenerate ground states~\cite{alicea2016topological}. Further, there exists a strong zero-mode~\cite{fendley2012parafermionic} that connects states in different fermionic parity sectors throughout the spectrum  and was found by Fendley~\cite{fendley2016strong}. In our case it can be written explicitly upon performing non-local transformation Eq.\eqref{nlt} as
\begin{equation}
    \Psi = \sum_{b=1}^\infty\sum_{S = \{a\}} \Delta^{2 - 2b} \left[ \prod_{j < b} (1 - 2 c_j^\dagger c_j) \right] (c_b^\dagger + c_b) \prod_{s = 1}^{S} \mathcal{F}_{a_{2s - 1} a_{2s}},
    \label{strongzeromode}
\end{equation}
where the sum runs over all sets $ S = \{a_1, a_2, \dots, a_{2S}, b\} $, with the positions ordered such that $ a_S < b $. The product is defined to be $1$ when $ S = 0 $ and the operator $\mathcal{F}_{aa'} $ is given by Fendley as
\begin{align*}
\mathcal{F}_{aa'}  &= \frac{1 - \Delta^2}{\Delta^{a - a'}} \Bigg[ 
(1 - 2 c^\dagger_a c_a)(1 - c^\dagger_{a'} c_{a'}) \\
&\quad - (c_a - c_a^\dagger)\, e^{i\pi \sum_{n=a}^{a'-1} c_n^\dagger c_n}\, (c_{a'} - c_{a'}^\dagger)
\Bigg].
\end{align*}
Notice that when $b=0$, the exact strong zero mode reduces to the weak zero mode $c_1+c^\dagger$, which is restricted to the ground state subspaces~\cite{wada2021coexistence}. 
\begin{figure}
    \centering
    \includegraphics[width=1\linewidth]{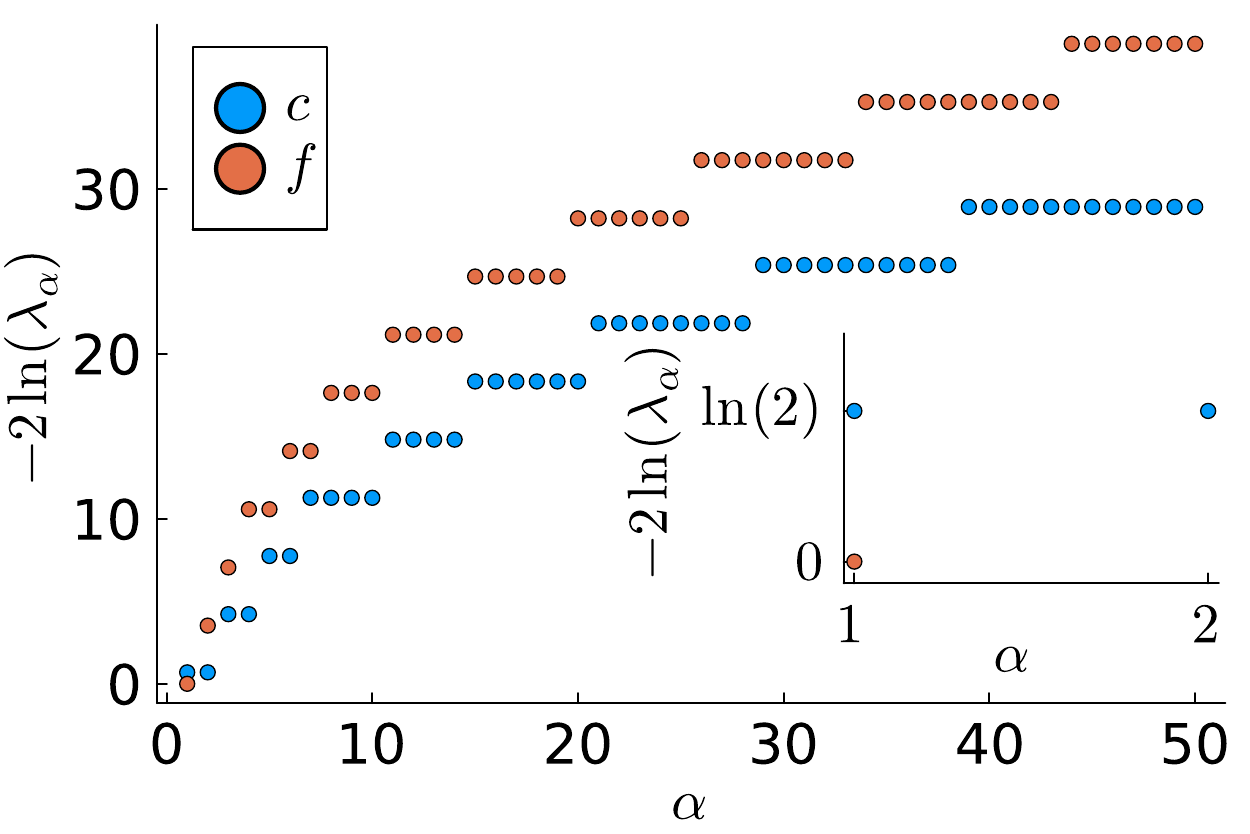}
    \caption{Entanglement spectrum of the Hamiltonian Eq.\eqref{modelham} and Hamiltonian Eq.\eqref{trivham} for $\Delta = 3$ and $N = 1000$, computed via DMRG using the ITensor library with a truncation cutoff of $10^{-16}$, are shown in blue and red color, respectively. For finite values of $\Delta>1$, the entanglement spectrum of Hamiltonian Eq.\eqref{modelham} exhibits an exact even degeneracy at every level, whereas, for Eq.\eqref{trivham}, the degeneracy could be even or odd. This perfect even degeneracy in entanglement for Hamiltonian Eq.\eqref{modelham} is evident in the representative set of the 50 lowest-lying levels and is indicative of nontrivial topological order in the system. In the inset, we show the exact result obtained for $\Delta<-1$, where the entanglement spectrum is exactly twofold degenerate for Hamiltonian Eq.\eqref{modelham} while it vanishes for the SSB Hamiltonian Eq.\eqref{trivham}.}
    \label{fig:EEdelp}
\end{figure}
We now turn to the $\Delta > 1$ regime, where, unlike the $\Delta < 1$ case, the ground state lacks simplicity. Using DMRG calculations implemented in the ITensor library, we compute the entanglement spectrum for a representative case with $\Delta = 3$ and $N = 1000$, setting a truncation cutoff of $10^{-16}$. The lowest $50$ levels, shown in Fig.~\ref{fig:EEdelp}, display perfectly even degeneracy at each level, a signature of the nontrivial topological phase that persists across the entire $\Delta > 1$ regime.

Moreover, we can construct a non-local string order parameter 
\begin{equation}
    \mathcal{O}^2_A=\lim_{|i-j|\to \infty}(-1)^{i-j}(c_i^\dagger + c_i)\,\exp\left(i\pi\sum_{n=i}^{j-1} c_n^\dagger c_n\right)\,(c_j^\dagger + c_j),
\end{equation}
whose explicit computation is simpler in the spin basis upon performing the inverse transformation Eq.\eqref{nlt} such that
\begin{equation}
    \mathcal{O_A}=\sqrt{\langle \mathcal O^2_A \rangle}=\lim_{|i-j|\to \infty}\sqrt{(-1)^{i-j}\langle \sigma_i^z \sigma_j^z \rangle} = \sigma,
    \label{OAdef}
\end{equation}
where $\sigma$ is staggered magnetized of the XXZ chain in $\Delta>1$ regime given by Eq.~\eqref{staggeredmag}. As illustrated in Fig.\ref{fig:pd}, the string order parameter increases monotonically from 0 at $\Delta = 1$ and asymptotically approaches 1 as $\Delta \to \infty$.  
In this regime, the exact strong zero mode is given by Eq.\eqref{strongzeromode}.

In summary, we have identified the symmetry-protected topological–spontaneous symmetry breaking (SPT‑SSB) correspondence for the paradigmatic XXZ model. Our key result is that the ferromagnetic and antiferromagnetic gapped phases—each characterized by spontaneous symmetry breaking—are mapped by our modified Jordan–Wigner transform into distinct symmetry-protected topological phases, whereas the intervening critical paramagnetic phase remains gapless and is mapped to a topologically trivial regime. Within each topological regime, a robust Majorana zero mode emerges, its wave function located exponentially at the boundary.

Acknowledgment: The authors thank Adrian B. Culver, Colin Rylands, and J. H. Pixley for their careful reading of the manuscript and insightful comments.
\bibliography{ref}

\widetext

\appendix

\section{The XXZ model - Summary of results }

We  summarize some of the well known results of the model. When the anisotropy parameter $\Delta>1$, the model has a spontaneously broken $\mathbb{Z}_2$ symmetry generated by  
$
\Pi = \prod_i \sigma^x_i
$ generating an antiferromagnetic order~\cite{syljuaasen2003entanglement}, and a unbroken $U(1)$ symmetry generated by $S^z=\sum_{i}\sigma^z_i$ ~\cite{takahashi1999thermodynamics}. In the thermodynamic limit and the model has mass gap~\cite{franchini2017introduction} $m=2\sinh\eta \frac{\vartheta _1^{\prime }\left(0,e^{-\eta }\right) \vartheta _3\left(\frac{\pi }{2},e^{-\eta }\right)}{ \vartheta _2\left(0,e^{-\eta }\right) \vartheta _4\left(\frac{\pi }{2},e^{-\eta }\right)}$ where $\vartheta_\alpha(u,q)$ for $\alpha=1,2,3,4$ are Jacobi theta functions, prime denotes the derivative with respect to the $u$ variable~\cite{kattel2024edge} and $\eta=\cosh^{-1}(\Delta)$. The staggered magnetization characterizing the symmetry breaking is given by~\cite{baxter1973spontaneous,izergin1999spontaneous} 
\begin{equation}
    \sigma = \lim_{N \to \infty} \frac{1}{N} \sum_i (-1)^i \langle \sigma^z_i \rangle=\frac{4 \left(e^{-2 \eta };e^{-2 \eta }\right)_{\infty
   }^2}{\left(-1;e^{-2 \eta }\right)_{\infty }^2}
   \label{staggeredmag}
\end{equation}
where $(a;q)_\infty = \prod_{k=0}^\infty (1 - a q^k)$ is the q-Pochhammer symbol~\cite{weisstein2008q}.

When $\Delta=1$, the model is gapless and its IR description is provided by $SU(2)_1$ Wess-Zumino-Witten conformal field theory~\cite{affleck1987critical} whereas when $\Delta = -1$, the spin-$1/2$ XXZ chain is in a gapless ferromagnetic phase with $N + 1$-fold degenerate ground state manifold corresponding to the total spin-$S = N/2$ multiple with magnon dispersion that is quadratic near zero momentum and hence is not described by a conformal field theory~\cite{franchini2017introduction}. These states are connected by global spin rotations. For $ -1 < \Delta < 1 $, the XXZ chain is in a gapless critical phase whose long-distance behavior is described by a compactified free boson conformal field theory with central charge $ c = 1 $ and Luttinger parameter $ K $ (related to $ \Delta $ via $ 2K = \frac{\pi}{\pi - \arccos(\Delta)} $~\cite{affleck1988field,giamarchi2003quantum}. Correlation functions exhibit power-law decay with exponents determined by $ K $, and no spontaneous symmetry breaking occurs.

And for $\Delta<-1$, the ground state is two-fold degenerate with $S=\pm \frac{N}{2}$.  The lowest excitations are gapped magnons—single spin flips that reduce the total magnetization by 1—with a gap given by $m = 4(|\Delta| - 1)$. In the thermodynamic limit, the system selects a symmetry-breaking ground state, and the gap protects long-range ferromagnetic order.

\section{Non-local unitary transformation}\label{dets}
Consider the anisotropic Heisenberg spin chain Hamiltonian

\begin{align}
H_{XXZ}
&=\sum_{i=1}^{N-1}\Bigl(\sigma^x_i\sigma^x_{i+1}+\sigma^y_i\sigma^y_{i+1}+\Delta\,\sigma^z_i\sigma^z_{i+1}\Bigr)
\label{xxzham}
\end{align}

Introduce a non-local unitary transformation to write the model in terms of the fermionic variable
\begin{equation}
    d_i^\dagger = U(\theta)\,\Bigl[\frac12\!\bigl(\prod_{j<i}\sigma^z_j\bigr)\,(\sigma^x_i+i\sigma^y_i)\Bigr]\,U^\dagger(\theta),
\end{equation}
with
\begin{equation}
    U(\theta)=
\begin{pmatrix}
\cos\frac{\theta}{3} + \tfrac{i}{\sqrt3}\sin\frac{\theta}{3}
& \tfrac{1+i}{\sqrt3}\sin\frac{\theta}{3}\\[6pt]
-\tfrac{1-i}{\sqrt3}\sin\frac{\theta}{3}
& \cos\frac{\theta}{3} - \tfrac{i}{\sqrt3}\sin\frac{\theta}{3}
\end{pmatrix},
\end{equation}
such that $U(\theta)U^\dagger(\theta)=\mathbb{I}$ for $\theta \in (0,2\pi)$. One can immediately recognize that for $\theta=0$, this is the usual Jordan-Wigner transformation, and for $\theta=2\pi$, it is the transformation we used in the main text. 

After this non-local transformation, the Hamiltonian becomes

\begin{align}
H(\theta)=\sum_{i=1}^{N-1}\Bigl\{&
\bigl[2+(\Delta-1)(\alpha^2+\beta^2)\bigr]\,(d^\dagger_{i+1}d_i + d^\dagger_i d_{i+1})\\
&+\bigl[1+(\Delta-1)\gamma^2\bigr]\,(2n_i^d-1)(2n_{i+1}^d-1)\\
&+(\Delta-1)(\alpha^2-\beta^2)\,(d_i^\dagger d_{i+1}^\dagger + d_i d_{i+1})\\
&+(\Delta-1)\,i\,\beta\gamma\;\prod_{j=1}^{i-1}(1-2n_j^d)\;\bigl[(d_i-d_i^\dagger)(1-2n_{i+1}^d)+(d_{i+1}-d_{i+1}^\dagger)\bigr]\\
&+(\Delta-1)\,\gamma\alpha\;\prod_{j=1}^{i-1}(1-2n_j^d)\;\bigl[(d_{i+1}+d_{i+1}^\dagger)+(d_i+d_i^\dagger)(1-2n_{i+1}^d)\bigr]
\Bigr\},
\end{align}

where

\begin{align}
\alpha(\theta)&=\tfrac{1}{3}\bigl(-\sqrt3\,\sin\tfrac{2\theta}{3}-\cos\tfrac{2\theta}{3}+1\bigr),\\
\beta(\theta)&=\tfrac{1}{3}\bigl(\sqrt3\,\sin\tfrac{2\theta}{3}-\cos\tfrac{2\theta}{3}+1\bigr),\\
\gamma(\theta)&=\tfrac{1}{3}\bigl(2\cos\tfrac{2\theta}{3}+1\bigr).
\end{align}

In the two limiting cases $\theta = 0$ and $\theta = 2\pi$, we obtain local Hamiltonians with specific parameter values. For $\theta = 0$, the parameters take the values $\alpha = 0$, $\beta = 0$, and $\gamma = 1$, yielding the Hamiltonian
\begin{equation}
H(0) = \sum_{i=1}^{N-1} \left[2\left(d^\dagger_{i+1} d_i + d^\dagger_i d_{i+1}\right) + \Delta\,(2n_i^d - 1)(2n_{i+1}^d - 1)\right].
\end{equation}
In contrast, for $\theta = 2\pi$, we have $\alpha = 1$, $\beta = 0$, and $\gamma = 0$, leading to the Hamiltonian
\begin{equation}
H(2\pi) = \sum_{i=1}^{N-1} \left[(1+\Delta)\left(d^\dagger_{i+1} d_i + d^\dagger_i d_{i+1}\right) + (1 - \Delta)\left(d_i d_{i+1} + d^\dagger_{i+1} d^\dagger_i\right) + (2n_i^d - 1)(2n_{i+1}^d - 1)\right].
\end{equation}

When $\theta=0$,  the transformation is the well-known Jordan-Wigner transformation, and the resultant local fermionic model is in a trivial charge-density-wave insulator. In contrast, when $\theta=2\pi$, the transformation is the new non-local transformation defined in the main text, and the resultant local fermionic model is in a non-trivial SPT phase.

Since the non-local $H(\theta)$ and $H_{XXZ}$ are related by a unitary rotation for every $\theta$, the entire spectrum—and in particular the gap—remains exactly constant as $\theta$ is varied from $0$ to $2\pi$ (for $\Delta>1$). Thus, in this non-local Hamiltonian $H(\theta)$, it is possible to change the parameter $\theta$ smoothly while maintaining the gap constant and to go from a trivial insulator to a non-trivial SPT order. However, we stress that this does not conflict with the standard assertion that a trivial insulator and an SPT phase cannot be connected without closing the gap.  That assertion presumes the interpolation is generated by a strictly local Hamiltonian (or, equivalently, by a finite‑depth quantum circuit).  In our construction, however, the family $U(\theta)$ is inherently non‑local—equivalent to an infinite‑depth circuit—and thus falls outside the usual locality constraints.  Consequently, the adiabatic continuity we exhibit is fully compatible with the conventional classification: under any finite‑depth, local evolution, the two phases remain distinct, yet they become unitarily connected once non‑local rotations are permitted.

\end{document}